\documentclass[twocolumn]{aa}
\usepackage{graphicx}
\usepackage{txfonts}
\begin{document}

\title{A possible signature for quark deconfinement 
           in  the compact star in  4U 1728-34}

  \author{Ignazio Bombaci 
          \inst{}
          }


   \institute{Dipartimento di Fisica ``Enrico Fermi'', Universit\'a di Pisa and 
     INFN Sezione di Pisa, via Buonarroti   2, I-56127 Pisa, Italy \\
              \email{bombaci@df.unipi.it}
             }

\date{Received ..............;  accepted ...................}

   \abstract{
   In a very recent paper Shaposhnikov {\it et al.}  (\cite{shapo}) 
   have extracted very tight constraints on the radius and the mass for the compact 
   star in the low mass X-ray binary 4U~1728-34,  from the analysis of a set 
   of Type I X-ray bursts  from this source.   
   In the present  letter, we  perform  a systematic comparison between the 
   mass-radius (MR) relation given by Shaposhnikov {\it et al.},  with the theoretical 
   determination of MR curves  for compact stars using  some of the most 
   recent and realistic models for the equation of state for stellar dense matter.  
  Our study clearly reveals that the semi-empirical MR relation for the compact 
  star in 4U~1728-34 is not compatible with models of neutrons stars 
  composed of nuclear matter or hyperonic matter, while it is consistent 
  with strange stars or neutron stars with a core of deconfined quark matter 
  (hybrid neutron stars).

   }

   \maketitle
%

\section{Introduction}

An accurate measure of the radius and the mass of an individual 
``neutron star'' will represent an extraordinary physical information to solve 
the long-standing puzzle on the internal constitution of these fascinating 
astrophysical bodies and to discriminate between different models 
for the equation of state (EOS) of dense hadronic matter. 

In a very recent paper Shaposhnikov {\it et al.}  (\cite{shapo}) (hereafter STH) 
have analyzed a set of 26 Type-I X-ray bursts  for the low mass X-ray binary 
4U~1728-34. The data were collected by the Proportional Counter Array 
on board of the Rossi  X-ray Timing Explorer (RXTE) satellite. 
For the interpretation of these observational  data 
Shaposhnikov {\it et al.}  (\cite{shapo})  used a model of the X-ray burst 
spectral formation developed by Titarchuk (\cite{titarchuk94}) and  
Shaposhnikov \& Titarchuk (\cite{ST02}).  
Within this model, STH were able to extract very stringent constrain on 
the radius and the mass of the compact star in this bursting source.  

In the present  letter, we  perform  a systematic comparison between the 
MR relation given by STH  with the theoretical determination of MR curves  
for compact stars.  We have calculated the MR curves for non-rotating compact 
stars in general relativity using some 
of the most recent and realistic models for the equation of state for stellar 
dense matter.  In particular, we consider the possibility that the compact star 
in 4U~1728-34 could have a core of deconfined quark matter
 (hybrid neutron star) or the possibility it could be a  strange star.

\section{Results}
The radius and mass for 4U~1728-34, extracted by STH for different best-fits 
of the burst data, are depicted in Figure~1 by the filled circles. 
Each of the four MR  points is relative to a different value of the distance to 
the source (d = 4.0, 4.25, 4.50, 4.75 kpc, for the fit which produces the smallest 
values of the mass, up to the one which gives the largest mass).   
The error bars on each point represent the error contour for 90\% confidence level.  

An additional restriction on the possible values of the radius and mass 
of the compact object in 4U~1728-34, has been  derived 
by Li {\it et al.}  (\cite{li99}), using the transition layer model  
(Titarchuk \& Osherovich \cite{to99}) to fit  the observed Quasi Periodic 
Oscillations (QPOs) in the persistent emission of this source. 
The stellar radius $R$ must be smaller than the inner radius $R_0$  
for the accretion disk around 4U~1728-34, which is given by 
$R_0 = 9 (M/M_\odot)^{1/3} {\rm km}$   (Titarchuk \& Osherovich \cite{to99}, 
Li {\it et al.} \cite{li99}). 
In addition, $R_0$ must be larger than the radius $R_{ms}$  of the last 
stable circular orbit around the star,   plotted as a dotted  curve 
in Figure 1 (Li {\it et al.} \cite{li99}).  
Therefore, according to Li {\it et al.} (\cite{li99})  
(see also Bombaci {\it  et al.} \cite{btd00}),  the allowed range of the 
mass and radius for 4U~1728-34 is the region in the lower corner of the 
MR plane confined by the dashed curve ($R= R_0$) and 
by  the dotted curve.  

It is very encouraging to notice that the analysis of two very different 
astrophysical phenomena (QPOs and X-ray bursts)  associated to this source,  
produces constraints on the radius and mass of the compact star which are 
consistent each other. 
  
Next we consider the theoretical mass-radius curves calculated 
solving the stellar structure equations in general relativity for 
non-rotating stars and using a representative set for  some of the most 
recent and realistic models for the equation of state for dense stellar matter.     
The three  curves labeled  BBB1, BPAL32 and GM3,  represent  
the MR relation for ``conventional'' neutron stars,  
{\it i.e.} for stars whose  core  is assumed to be composed by an 
uncharged mixture of neutrons, protons, electrons and muons in 
equilibrium with respect to the weak interaction.    
The curve BBB1 refers to the stellar calculations based on the microscopic EOS 
for $\beta$-stable nuclear matter computed by Baldo {\it et al.} (\cite{bbb}) 
using the Brueckner-Bethe-Goldstone many-body theory with realistic 
two-body plus three-body nuclear interaction.     
The curve BPAL32  shows the MR relation calculated with a phenomenological EOS 
for nuclear matter (Bombaci \cite{bomb95}, Prakash {\it et al.} \cite{prak97})  
derived by  a density-dependent effective NN interaction. 
The curve GM3 is relative to a neutron star sequence calculated within 
a relativistic field theoretical approach in the mean field approximation 
(see {\it e.g.} Serot \& Walecka \cite{wale86});   here in particular,  
we used the GM3 parametrization (for the pure nucleonic case) 
given by Glendenning and Moszkowski (\cite{GM}).   
Other realistic models for the EOS of nuclear matter, as for example the one  
of Wiringa {\it et al.}  (\cite{wff}) or that of Akmal  {\it et al.}  (\cite{apr98}),  produce 
neutron stars with ``large'' radii between 10 -- 13 km 
(in the mass range 0.8 $M_\odot$ -- $M_{max}$).  

The curve labeled Hyper depicts the MR relation for a neutron star in which 
hyperons are considered in addition to nucleons as hadronic constituents. 
Here we used the parameter set GM3 of reference (Glendenning and Moszkowski \cite{GM}).   
The MR curve labeled $K^-$ is relative to neutron stars   
(Glendenning \& Schaffner-Bielich  \cite{gs99})   
with a Bose-Einstein condensate of negative kaons in their cores. 

It is clearly seen in Figure~1 that none of the neutron star MR curves,   
for all the EOS models described so far,  is consistent with the radius and the 
mass for 4U~1728-34 extracted by  STH.  
Therefore  4U~1728-34 is not well described by an {\it hadronic star}  model 
({\it i.e.} a neutron star with a core made of different hadron species).      

Next we consider the possibility the compact star in 4U~1728-34 is 
a {\it strange star}, {\it i.e.} a compact star  consisting completely 
of a deconfined mixture of {\it up} ({\it u}),  {\it down} ({\it d})  and  
{\it strange } ({\it s}) quarks  (together with an appropriate number of electrons 
to guarantee electrical neutrality)  satisfying the Bodmer--Witten hypothesis 
(Bodmer \cite{bod71}, Witten \cite{wit84}).      
The two curves  labeled B70 and  B85 in Figure~1 give the MR relation for 
strange stars described by an EOS  for strange quark matter 
(Farhi \& Jaffe \cite{fj84}) based on the MIT bag model for hadrons. 
These  two MR curves are obtained taking  B = 70 MeV/fm$^3$, 
and  B = 85 MeV/fm$^3$.   
In both cases the mass of the strange quark is $m_s = 150$~MeV,  
while {\it up} and {\it down} quarks are considered massless.   
The curve SS1  gives the MR relation for strange stars calculated 
with the EOS by Dey {\it et al.} (\cite{dey98}).       

Finally, we consider the possibility the compact star in 4U~1728-34 is 
an {\it hybrid neutron star}, {\it i.e.} a compact star which possess a quark matter 
core either as a mixed phase of deconfined quarks and hadrons, 
or as a pure quark matter phase.  
In Figure~1, we plot the MR curves obtained for hybrid stars  
with the GM3 equation of state for the hadronic phase (case with hyperons) 
(Glendenning and Moszkowski \cite{GM}), and with the  bag model EOS for the 
quark phase (Farhi \& Jaffe \cite{fj84}) taking  
B =   80~MeV/fm$^3$, $m_s= 150$~MeV, $m_u = m_d = 0$ (curve Hy1), or with 
B = 100~MeV/fm$^3$, $m_s = m_u = m_d = 0$  (curve Hy2).   

Figure~1 clearly demonstrates that a strange star or an hybrid star model 
is more  compatible with 4U~1728-34  than a neutron star one.   
Here we have considered a limited yet representative set for some of the most 
recent models for the equation of state for dense stellar matter.       
We have checked that other models for hybrid stars and strange stars 
are in agreement with the MR for 4U~1728-34 extracted by STH. 
This is particularly true, for example, for some of the hybrid star models 
discussed by Burgio {\it et al.} (\cite{bbss02}) where a density dependence of the 
bag constant is introduced to describe the quark phase, 
or in the case of strange stars  (Drago  \& Lavagno \cite{dl01})  calculated 
within the Color Dielectric Model for strange quark matter.  
 
\section{Conclusions}

Our study clearly reveals that the semi-empirical MR relation for the compact 
star in 4U~1728-34 obtained by STH  is not compatible with models of neutrons stars 
with cores composed of nuclear matter or hyperonic matter ({\it hadronic stars}), 
while it is consistent with strange stars or neutron stars with a core of deconfined 
quark matter (hybrid neutron stars).



   \begin{figure*}
~
  \vskip 2.0cm 
   \centering
\includegraphics[width=16cm]{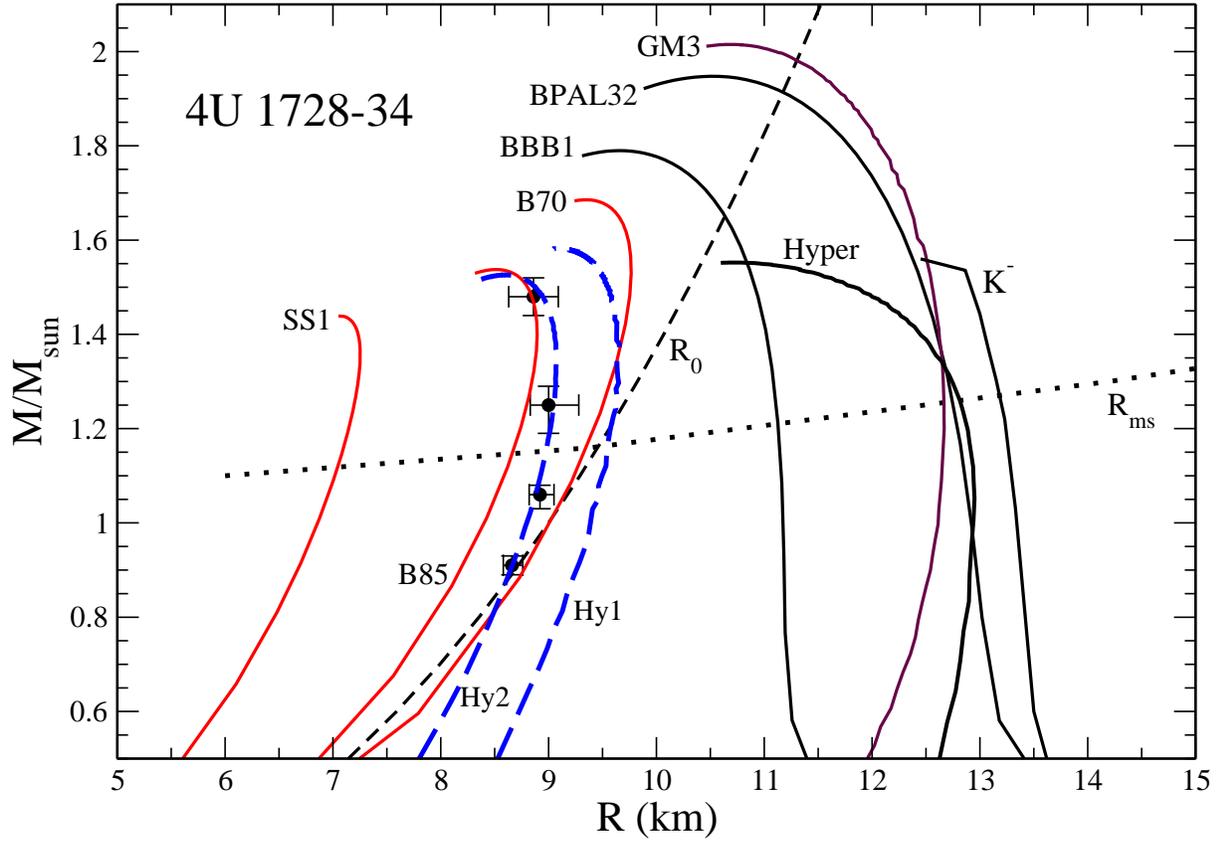}
\vskip 0.5cm
   \caption{ The radius and mass for 4U~1728-34, extracted by 
Shaposhnikov {\it et al.}  (\cite{shapo})  for different best-fits 
of the X-ray burst data, is shown by the filled circles.   
The error bars on each point represent the error contour for 90\% confidence level.  
We show also the restrictions on the possible values of the radius and mass 
for 4U~1728-34  derived by Li {\it et al.}  (\cite{li99})   
(region in the lower corner of the MR plane confined by the dashed curve ($R= R_0$)  
and by  the dotted curve).    
The three  curves labeled  BBB1, BPAL32 and GM3,  reprersent 
the MR relation for ``conventional'' neutron stars. 
The curve labeled Hyper shows the MR relation for a neutron star 
with an hyperonic core, and the curve  labeled $K^-$ is relative to neutron stars    
with kaon condensation.  
The two curves Hy1 and Hy2   give the MR relation for hybrid stars. 
Finally, the curves B70, B85, and SS1 are relative to strange stars. 
} 

              \label{fig}%
    \end{figure*}
\end{document}